\documentclass[times,twocolumn,review]{elsarticle}

\usepackage{medima}
\usepackage{framed,multirow}

\usepackage{amssymb}
\usepackage{latexsym}

\usepackage{url}
\usepackage{xcolor}

\usepackage{hyperref}

\usepackage{amsmath}
\usepackage{algorithmic}
\usepackage{graphicx}
\usepackage{url}
\usepackage{multirow}
\usepackage{tabularx}

\definecolor{newcolor}{rgb}{.8,.349,.1}

\journal{Medical Image Analysis}

\begin{document}

\verso{Given-name Surname \textit{et~al.}}

\begin{frontmatter}

\title{Brain Connectivity Network Structure Learning For Brain Disorder Diagnosis}%


\author[mymainaddress]{Dongdong~Chen}
\author[mymainaddress]{Linlin~Yao}
\author[mymainaddress]{Mengjun~Liu}
\author[mymainaddress]{Zhenrong~Shen}
\author[mymainaddress]{Yuqi~Hu}
\author[mymainaddress]{Zhiyun~Song}
\author[mysecondaddress]{Shengyu~Lu}
\author[mythirdaddress]{Qian~Wang}
\author[mythirdaddress]{Dinggang~Shen}
\author[mymainaddress]{Lichi~Zhang*}

\cortext[mycorrespondingauthor]{Corresponding author: lichizhang@sjtu.edu.cn (Lichi Zhang)}




\address[mymainaddress]{School of Biomedical Engineering, Shanghai Jiao Tong University, Shanghai 20030, China}
\address[mysecondaddress]{College of Computer Science and Technology, Harbin Engineering University, Harbin 150001, China}
\address[mythirdaddress]{School of Biomedical Engineering, ShanghaiTech University, Shanghai 201210, China}


\begin{abstract}
Recent studies in neuroscience highlight the significant potential of brain connectivity networks, which are commonly constructed from functional magnetic resonance imaging (fMRI) data for brain disorder diagnosis.
Traditional brain connectivity networks are typically obtained using predefined methods that incorporate manually-set thresholds to estimate inter-regional relationships. However, such approaches often introduce redundant connections or overlook essential interactions, compromising the value of the constructed networks. Besides, the insufficiency of labeled data further increases the difficulty of learning generalized representations of intrinsic brain characteristics.
To mitigate those issues, we propose a self-supervised framework to learn an optimal structure and representation for brain connectivity networks, focusing on individualized generation and optimization in an unsupervised manner. We firstly employ two existing whole-brain connectomes to adaptively construct their complementary brain network structure learner, and then introduce a multi-state graph-based encoder with a joint iterative learning strategy to simultaneously optimize both the generated network structure and its representation. By leveraging self-supervised pretraining on large-scale unlabeled brain connectivity data, our framework enables the brain connectivity network learner to generalize effectively to unseen disorders, while requiring only minimal fine-tuning of the encoder for adaptation to new diagnostic tasks.
Extensive experiments on cross-dataset brain disorder diagnosis demonstrate that our method consistently outperforms state-of-the-art approaches, validating its effectiveness and generalizability. The code is publicly available at \url{https://github.com/neochen1/BCNSL}.
\end{abstract}


\end{frontmatter}


\section{Introduction}
\label{sec1}

Resting-state functional magnetic resonance imaging (rs-fMRI) is a non-invasive imaging technique that has demonstrated its significant potential for evaluating brain activity by measuring blood oxygen level-dependent (BOLD) signals over time~\cite{wee2016fmri}. 
Researchers employ rs-fMRI time series to investigate brain connectivity networks, which can explore the mechanisms of brain function and information flow, thereby contributing to the diagnosis of brain disorders~\cite{hallett2020human}.





\begin{figure*}[!t]
	\begin{center}
		{\centering\includegraphics[width=0.9\linewidth]{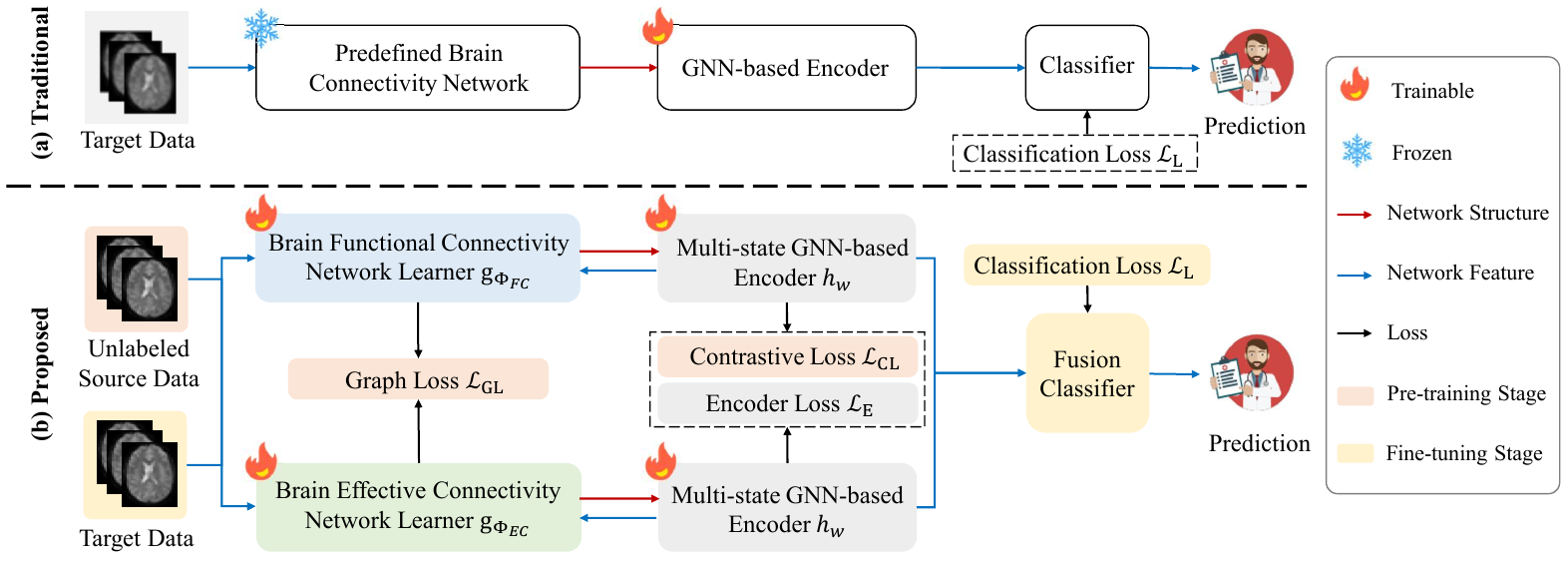}}
		\caption{A sketch of the proposed brain connectivity network structure learning framework compared with the traditional method.}
	    \label{sketch}
	\end{center}
\end{figure*}

Current methods for analyzing brain connectivity networks typically follow a two-stage paradigm as shown in Fig.~\ref{sketch} (a): (1) constructing the brain connectivity network using predefined methods, and (2) extracting brain network representations for disorder diagnosis. For the first stage, the brain connectivity network is typically modeled as a graph structure, where each node denotes a brain region, and the edges indicate the strength of connections between these regions. Depending on the different types of connections, brain connectivity networks can generally be categorized into two types: functional connectivity (FC) and effective connectivity (EC)~\cite{farahani2019application}. FC describes statistical dependencies among brain regions, while EC characterizes the causal effects from one brain region to another~\cite{mukta2017theory,nag2023dynamic}. Furthermore, integrating these two distinct patterns can provide comprehensive insight into the analysis of brain connectivity networks. For the second stage, methods based on graph neural networks (GNNs) have been widely used as an encoder to extract representations of brain connectivity networks, due to their powerful ability to extract valuable topological information in graph structure~\cite{xu2018how}.

The main challenge for the existing brain connectivity network methods is that they are generally trained and applied to specific and limited brain disease diagnosis tasks. Therefore, those trained models cannot be directly employed for other brain disorders, but instead need to be retrained from scratch to achieve satisfactory results. Such generalization issue has hindered their practical applications in clinical diagnosis, since it is generally high cost to collect sufficient number of samples of specific brain disorder disease for model training.   

Many studies attribute the aforementioned issue to the insufficient feature extraction ability of models, and focus on designing powerful GNN-based brain network encoders that are eligible to acquire a universal representation of the brain connectivity network across various brain disorders. 
For example, Wang~\textit{et al.}~\cite{wang2023unsupervised} proposed two parallel spatio-temporal graph convolutional neural networks for fMRI feature extraction. Chen~\textit{et al.}~\cite{10735236} designed a heterogeneous graph convolutional neural network model with the aim of capturing robust brain network embeddings.

\begin{figure}[!t]
	\begin{center}
		{\centering\includegraphics[width=1.0\linewidth]{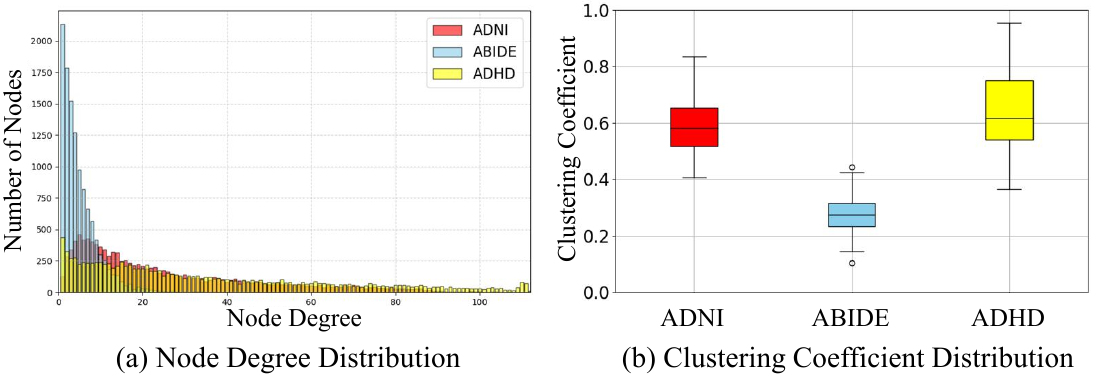}}
		\caption{An exemplar case to show the distribution of brain functional connectivity networks constructed using predefined methods with 100 subjects randomly selected from the ADNI~\cite{adnidataset}, ABIDE~\cite{craddock2013neuro}, and ADHD~\cite{adhd2012adhd} datasets. The predefined brain connectivity networks are constructed using Pearson correlation with a threshold of 0.5 referring to previous works~\cite{wang2014systematic, david2004evaluation}. (a) is the distribution of node degrees, which reveals the overall connectivity pattern of the network. (b) is the distribution of clustering coefficients, which reflects the tightness of local structures in the network.}
	    \label{FC_distribution}
	\end{center}
\end{figure}

However, those methods fail to address the intrinsic properties of the input brain connectivity network, and the quality of the brain connectivity network graph structure plays a crucial role in the learning of brain networks. Although many works have focused on improving the quality of the graph structure for brain connectivity network representation learning, there are several limitations that hindered their model performance: 1) The predefined approaches heavily depend on prior knowledge and inevitably introduce noise. For example, some studies directly employ a fully connected graph constructed using Pearson correlation, which obviously includes many redundant connections~\cite{cui2022braingb}. Some researchers manually set a threshold value to control the sparsity of the graph, which potentially results in the loss of key information~\cite{wang2014systematic, li2021braingnn}. 2) Due to individual differences and heterogeneity in data collection, fixed brain network structures constructed using predefined methods exhibit significant distribution differences across different datasets, which increases the difficulty of cross-dataset analysis for the model. As shown in Fig.~\ref{FC_distribution}, the degree of nodes of subjects in the ADHD dataset shows a wide distribution, while the graph structure in the ABIDE dataset exhibits a significantly low connectivity. 3) The predefined brain connectivity networks are invariant during the data preprocessing stage, which is independent of downstream brain network analysis models, making it theoretically difficult to achieve optimal performance in the entire framework.

Given the aforementioned issues, our target is to develop a framework to adaptively derive the optimal brain connectivity network for various diseases through structural learning, thereby enhancing the generalizability and performance of cross-disease diagnosis.
As illustrated in Fig.~\ref{sketch} (b), we propose two brain connectivity network learners, which automatically generate personalized connectivity patterns from distinct views. 
These learners are jointly optimized with the encoder within an end-to-end iterative learning framework, allowing the model to generate robust and high-quality structures and representations of the brain network.
To improve generalization to novel brain disorders with limited labeled data, we introduce a contrastive learning framework for pre-training the model using abundant unlabeled data. Once pre-trained, the model can be effectively adapted to new conditions by fine-tuning only the encoder, and multi-view fusion information is utilized to achieve effective diagnosis of new brain diseases.
The main contributions of this work can be summarized as follows:

\begin{itemize}
\item We propose novel brain connectivity network learners based on a multi-head similarity learning metric that can automatically generate different views of brain networks.

\item We integrate brain network construction and feature representation into a unified framework, and propose a joint iterative optimization strategy to simultaneously refine the graph structure and its corresponding representations.

\item The well-trained brain connectivity network learner can directly generate adaptive structures for new disorders without requiring any fine-tuning, which further reduces the model's dependence on labeled target samples.

\item We analyze the effect of several major components on the performance of the proposed method, and perform comprehensive experiments on four publicly available datasets to demonstrate its superiority.

\end{itemize}

Note that we have extended our preliminary work published at the MICCAI 2024 conference~\cite{chen2024self}, and we have made several major improvements to the effectiveness and generalizability of the brain connectivity network method for disorder diagnosis, which can further enhance its applicability in clinical practice: 
(1) We extend the simple similarity measurement among brain regions in~\cite{chen2024self} to a learnable multi-head feature space, which improves the stability and enhances the expressive power of the brain network structure learning process.
(2) To further enhance the framework’s capability in brain network structure modeling and feature learning, we propose a joint iterative optimization strategy based on the hypothesis that better node embeddings can be obtained from a better graph structure.
(3) Previous work in~\cite{chen2024self} has to be fine-tuned from scratch to achieve satisfactory results. Here, we enrich the methodology of brain network structure learning to enhance generalizability, allowing adaptation to new data with fine-tuning only the encoder.
(4) We have significantly expanded the datasets included for the pre-training and validation of the brain network analysis framework to better demonstrate the superiority of our method.


\section{Related Work}

\subsection{Graph Structure Learning for Brain Connectivity Network}

In brain network analysis, brain connectivity networks are typically predefined using a specific statistical method combined with an artificial threshold. However, this approach inevitably introduces noise or leads to incomplete brain connectivity, impeding a comprehensive understanding of the underlying disease mechanisms~\cite{he2010graph}. 
In pursuit of an optimal graph structure for brain connectivity networks, recent studies have sparked an effort around graph structure learning, which involves discovering and modeling relationships among complex variables as a graph automatically. 
For example, Kan~\textit{et al.}~\cite{kan2022fbnetgen} developed an end-to-end framework composed of a graph generator and predictor, which transform raw time-series features into task-oriented brain networks using the inner product. Febrinanto~\textit{et al.}~\cite{febrinanto2023balanced} proposed a balanced graph structure for brains, which utilizes graph convolutional layers to filter the original correlation matrix and generate an optimal sample graph.

However, most research on graph structure learning focuses on modeling a specific view of brain connectivity patterns (typically undirected graphs for brain functional connectivity), while neglecting the directional information flow among brain ROIs (i.e., directed graphs for brain effective connectivity). 
In addition, these methods rely on the supervision of downstream tasks, making them constrained by the need for large-scale labeled training data. Therefore, when faced with new brain disorders, they typically require training a new model for learning graph structures with extensive data, which poses significant challenges for clinical practice.





\subsection{Unsupervised Learning for Brain Disorder Diagnosis}

Unsupervised methods play a crucial role in promoting the progress of deep learning-based brain disorder diagnosis, which eliminates the dependence on large-scale labeled training data~\cite{naeem2023unsupervised}.
Existing unsupervised methods for brain disorder diagnosis are typically divided into two categories: 1) autoencoder-based framework, and 2) contrastive learning-based framework. In autoencoder-based frameworks, input data are first compressed into low-dimensional representations using encoders, and then reconstructed to recover the original input using decoders. This approach enables unsupervised feature extraction by minimizing reconstruction loss. Wen~\textit{et al.}~\cite{wen2023graph} proposed an ensemble masked graph method based on the autoencoder framework, which utilizes the partially visible nodes in the brain network as input and reconstructs the masked edges to learn latent brain network representations. However, random masking may result in the loss of information for critical brain ROIs, hindering the comprehensive capture of brain information.

In contrastive learning-based frameworks, the typical approach involves generating positive pairs of similar samples by augmenting data from the same sample, and constructing negative pairs of dissimilar samples based on different samples. The objective is to minimize the distance between positive pairs and maximize the distance between negative pairs within the embedding space. For instance, Wang~\textit{et al.}~\cite{wang2023unsupervised} introduced an unsupervised contrastive graph learning framework, which manually constructed pairs of samples for brain functional connectivity networks using two sliding windows from BOLD signals. Although contrastive learning frameworks are superior in capturing the inherent consistent structure of data, it is necessary to design positive and negative sample pairs reasonably with discriminative and semantic consistency of contrastive targets, thereby avoiding noise interference and promoting high-quality representation learning.

\begin{table}[!t]
	\centering
	\caption{Demographic information of the studied subjects from UKB, ADNI, ABIDE, and ADHD. NC; normal control, BD: brain disorder}
     \resizebox{\linewidth}{!}
     {
	\begin{tabular}{c c c c c c} 
  \hline
		Datasets & Number & Age Range & Age Statistics & Male/Female  &  NC/BD\\
		\hline
  UKB & 1499 & 46.0-79.0 & 60.54$\pm$7.38 & 757 / 742 & - \\
  ADNI & 114  & 61.0-90.0 & 72.58$\pm$6.05 & 58 / 56 & 60 / 54 \\
  ABIDE  & 182 & 6.47-39.10 & 15.21$\pm$6.62 & 145 / 37 & 104 / 78 \\
  ADHD & 257  & 7.17-17.96 & 11.59$\pm$2.89 & 165 / 92 & 110 / 147 \\

\hline

	\end{tabular}}
	\label{tab:dataset}
\end{table}

\section{Materials}
\subsection{Datasets}
In this study, we utilize four public datasets: UK Biobank (UKB\footnote{https://biobank.ctsu.ox.ac.uk/.})~\cite{sudlow2015uk}, Alzheimer's Disease Neuroimaging Initiative (ADNI\footnote{http://adni.loni.usc.edu/.})~\cite{adnidataset}, Autism Brain Imaging Data Exchange (ABIDE\footnote{http://preprocessed-connectomes-project.org/abide/.})~\cite{craddock2013neuro}, and Attention Deficit Hyperactivity Disorder (ADHD\footnote{http://preprocessed-connectomes-project.org/adhd200/.})~\cite{adhd2012adhd}. The demographic information of all the subjects included in our study is summarized in TABLE~\ref{tab:dataset}. It is worth noting that the subjects selected from UKB are unlabeled and are only used for unsupervised pre-training.



\subsection{Preprocessing}
For the UKB and ADNI datasets, we download the original rs-fMRI data. To preprocess the data for each subject, we first employ a standardized protocol provided by the Data Processing Assistant for Resting-State fMRI (DPARSF) toolbox. This includes performing slice timing correction, motion correction, spatial and temporal filtering, as well as covariate regression~\cite{processpipeline}. Subsequently, we utilize the anatomical automatic labeling (AAL) template to parcellate brain images. Finally, we obtain BOLD signals for each subject, which are depicted as a collection of sequences consisting of time points, and each sequence corresponds to a specific region of the brain.


For the ABIDE and ADHD datasets, we select the imaging center (New York University Medical Center) with the largest number of subjects, generally referring to the previous studies~\cite{xie2021constructing, sun2021estimating}. All rs-fMRI data used in our analysis are obtained from the Preprocessed Connectome Project Initiative, and the official provides results of preprocessing using various tools separately, including the Connectome Computation System (CCS), the Configurable Pipeline for the Analysis of Connectomes (CPAC), the DPARSF, and the NeuroImaging Analysis Kit. To maintain consistency with the UKB and ADNI datasets, we choose the same toolbox and strategies including filtering, regression, and segmentation. Consequently, we obtain the collection of BOLD signals for each sample in the ABIDE and ADHD datasets.


\section{Method}
\begin{figure*}[!t]
	\begin{center}
		{\centering\includegraphics[width=1.0\linewidth]{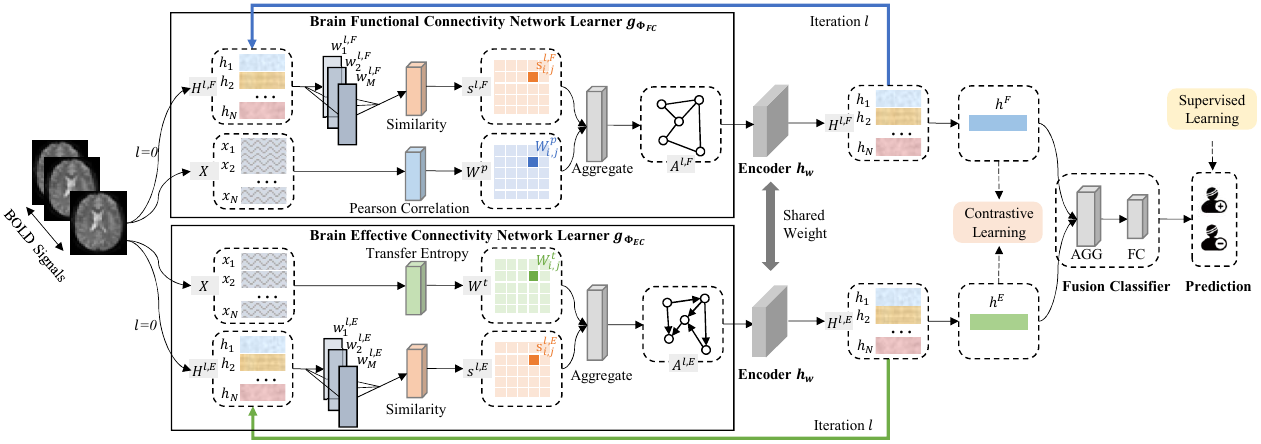}}
		\caption{Overview of the proposed method. First, the BOLD signals are used to construct learnable brain functional and effective connectivities. Then, the brain connectivity network encoder captures the hidden representation based on the generated brain connectivities. The representation is then fed into the brain connectivity network learner to optimize the learner and encoder iteratively. Finally, the outputs of both functional and effective connectivity embeddings are utilized for self-supervised pre-training and supervised downstream tasks.}
	    \label{framework}
	\end{center}
\end{figure*}

As shown in Fig.~\ref{framework}, a brain connectivity network learning framework is proposed for brain disorder diagnosis. This framework comprises two network learners that generate distinct brain connectivity patterns, along with corresponding encoders that capture network representations. The structure and representations are optimized jointly and iteratively, with a fused representation used to diagnose brain disorders. The whole framework is pre-trained using contrastive learning based on different brain connectivity features and then fine-tuned through supervised learning.

\subsection{Problem Formulation}\label{4a}
We define the diagnosis of different brain disorders with brain connectivity network learning as a cross-dataset classification task. The input datasets are composed of source data and target data denoted as $D=\{D_{source}, D_{target}\}$, where $D_{source}=\{X_{i}\}_{i=1}^P$, and $D_{target}=\{X_{i}, Y_{i}\}_{i=1}^Q$. Here, $X$ is the expression matrix of BOLD signals obtained from rs-fMRI data preprocessing, and $Y$ is the set of labels of all target subjects. 
All BOLD signals are defined using the AAL atlas and expressed as time sequences based on ROIs $X=\left(x_1, x_2, \cdots, x_N\right) \in \mathbb{R}^{N \times T}$, where $N$ is the number of ROIs and $T$ is the time points for BOLD signals. 
The model is pre-trained using the unlabeled source dataset and then fine-tuned on a randomly selected subset of labeled data from the target dataset. 
The final model predicts the results, denoted as $\hat{Y}$, for the remaining test subjects in the target dataset.

\subsection{Brain Connectivity Network Learner}\label{sec4b}
To overcome the limitations posed by the original predefined network structure, we propose autonomously acquiring intrinsic connections among different brain regions in a structure-learning manner. In contrast to the conventional fixed brain connectivity matrices defined by correlation and hand-crafted threshold, we directly generate two distinct brain connectivity networks from the BOLD signals. 


Specifically, for each subject that is expressed as a collection of BOLD signals $X=\left(x_1, x_2, \cdots, x_N\right) \in \mathbb{R}^{N \times T}$, our goal is to identify a generation function $A_{ij}=g_{\Phi}(x_{i},x_{j})$ that is parameterized to construct brain connectivity networks $A$ with each element defined as $A_{ij}$.

Compared to the brain connectivity learner in the previous version, which simply uses absolute subtraction, we design a novel weighted cosine similarity function. We first introduce a learnable weight vector $\mathbf{w}$ which has the same dimension as the input $x_i$ and $x_j$, and the initial weighted similarity metric among different brain regions can be formulated as

\begin{equation}
s_{i j}=\cos \left(\mathbf{w} \odot x_i, \mathbf{w} \odot x_j\right),
\end{equation}
where $\odot$ denotes the Hadamard product, and the learnable weight $\mathbf{w}$ learns to highlight different dimensions of the vectors.

To improve the stability and enhance the expressive power of the learning process, we further extend the learnable weight into multi-head space. In practice, $m$ weight vectors are employed in the above-mentioned similarity function to compute independent similarity matrices, and the average of these matrices is then considered as the final similarity.

\begin{equation}\label{eq:sij}
s_{i j}=\frac{1}{M} \sum_{M=1}^m \cos \left(\mathbf{w}_m \odot x_i, \mathbf{w}_m \odot x_j\right)
\end{equation}

Considering that there are two different patterns of brain network connectivity, i.e., brain functional connectivity networks (FCN) and effective connectivity networks (ECN) $A=\{A^{F}, A^{E}\}$, we design two implements $g_{\Phi}=\{g_{\Phi_{FC}}, g_{\Phi_{EC}}\}$ according to different interaction characteristics of the connections. Therefore, the weight $\mathbf{w}_m=\{\mathbf{w}^{F}_m, \mathbf{w}^{E}_m\}$ is learned in the functional connectivity leaner and effective connectivity learner, respectively.

In addition, we utilize Pearson correlation ${\rho}(x_{i}, x_{j})$~\cite{benesty2009pearson} and transfer entropy ${\zeta}(x_{i}, x_{j})$~\cite{vicente2011transfer} to measure the structural relationship $W_{ij}$ between paired ROIs, respectively, which can be formulated as

\begin{equation}
  W_{ij}=\left\{\begin{array}{l}
  W_{ij}^{p}=\rho\left(x_i, x_j\right), \text { if } g_{\Phi}=g_{\Phi_{FC}} \\
  W_{ij}^{t}=\zeta\left(x_i, x_j\right), \text { if } g_{\Phi}=g_{\Phi_{EC}}
  \end{array}\right.
\end{equation}


Finally, we integrate structural information with learnable relationships to generate brain connectivity networks as


\begin{equation}
A_{ij}=\frac{W_{ij} s_{ij}}  {\sum_{j=1}^n W_{ij} s_{ij}}.
\end{equation}

It is worth noting that, this formulation represents a general form of brain functional and effective connectivity, using different parameters (i.e., $\mathbf{w}_m$ and $W_{ij}$) depending on the connectivity pattern. Furthermore, due to the bidirectional and unequal process of information transmission (i.e., $A_{ij}=A_{ji}$, if $g_{\Phi}=g_{\Phi_{FC}}$, and $A_{ij} \neq A_{ji}$, if $g_{\Phi}=g_{\Phi_{EC}}$), the generated effective connectivity network based on the transfer entropy is represented as an asymmetric matrix, whereas the generated functional connectivity network based on the Pearson correlation is represented as a symmetric matrix.

\subsection{Joint Iterative Optimization}\label{4c}
After generating the brain connectivity network, it is essential to extract the brain information. Compared to traditional GNN models, this study employs a multi-state GNN-based encoder specifically designed to capture the brain’s multi-functional characteristics~\cite{chen2023learnable}. By incorporating different functional states $c$ of the brain, this encoder effectively maps the structure of the connectivity network to the representation matrix $H=\left(h_1, h_2, \cdots, h_N\right)$, where ${h}_{i}$ denotes the representation vector of each brain region. Therefore, the brain connectivity network encoder can be formulated as $H=\operatorname{Enocder}(A, c; \theta)$, where $c$ is hyperparameter and $\theta$ is learned by model training.

It is known that more refined node embeddings can be learned by employing an improved graph structure, meanwhile, a better graph structure can be acquired through improved node embeddings. Therefore, we further propose a joint iterative optimization strategy for the brain connectivity network learner and encoder.

Specifically, in the initial round (i.e., $l=0$) of brain connectivity network generation, both the learnable similarity relationship and structural information are derived based on the input of BOLD signals. In subsequent iterations, the learnable similarity is calculated based on the representation of brain connectivity networks obtained by encoders. Therefore, the formulation~\eqref{eq:sij} can be rephrased as:


\begin{equation}
s_{i j}^{l}=\frac{1}{M} \sum_{M=1}^m \cos \left(\mathbf{w}_m \odot h_{i}^{l}, \mathbf{w}_m \odot h_{j}^{l}\right), l \in [0,L].
\end{equation}
Here, the hidden representation of each brain region in the $l$-th iteration layer is denoted as ${h}_{i}^{l}$, and ${h}_{i}^{l}$ is set as $x_{i}$ when $l$ equals 0. $L$ is the number of iteration layers. In addition, for brain functional and effective connectivity networks, we utilize the feature vectors (i.e., ${h}_{i}^{l,F}$, and ${h}_{i}^{l,E}$) obtained from their respective encoders to calculate the corresponding similarity measures (i.e., $s_{i j}^{l,F}$, and $s_{i j}^{l,E}$).

We further aggregate the new structural information and similarity measures to obtain the brain connectivity network under each round of joint iterative optimization, which can be formulated as follows:

\begin{equation}
  A^{l}_{ij}=\frac{W_{ij} s_{i j}^{l}} {\sum_{j=1}^n W_{ij} s_{i j}^{l}}.
\end{equation}

It is important to note that the conventional structural information remains fixed throughout the entire iteration. Moreover, the brain connectivity network encoder is updated on the basis of the new brain connectivity matrices.

\begin{equation}
H^{l}=\operatorname{Enocder}(A^{l}, c; \theta).
\end{equation}
The $H^{l}=(h^{l}_{1},h^{l}_{2}, \cdots, h^{l}_{N})$ denotes the hidden representation matrix of brain connectivity networks in the $l$-th iteration. We further derive the representation vector for each brain connectivity network (i.e., $h=\{h^{F}, h^{E}\}$) with mean pooling operation $h=\frac{1}{N}H^{l}$, where the vector $h^{F}$ and $h^{E}$ are different results corresponding to the input of brain functional and effective connectivity matrix, respectively.

\subsection{Model Pre-training and Fine-tuning}\label{4d}
To pre-train the model, we first construct an unsupervised representation learning framework based on graph contrastive learning. Different from the traditional methods that use different time segments of brain connectivity networks as paired samples, we propose a comparative analysis of brain networks from different connectivity patterns.
Specifically, we consider the two types of generated brain connectivity networks (i.e., functional connectivity and effective connectivity) from the same subject input as the positive pair. The remaining networks from the same data batch $B$ serve as negative pairs, and then we design a contrastive loss based on normalized temperature-scaled cross-entropy loss (NT-Xent)\cite{you2022bringing}, which can be formulated as:

\begin{equation}
  \mathcal{L}_{CL}=-\frac{1}{B} \sum\nolimits_{i=1}^{B} \log \frac{\exp \left({h^{F}}_i \cdot {h^{E}}_i / \tau\right)}{\sum_{k=1, k\neq i}^{2B} \exp \left({h^{F}}_i \cdot {h}_k / \tau\right)},
\end{equation}
where $\tau$ is a temperature hyperparameter.


In addition, to optimize the structure of the learnable brain connectivity network in the final iteration layer $A ^ {L} $, we further propose a graph loss:

\begin{equation}
\mathcal {L}_ {\mathrm{GL}}=\sum\nolimits_{i, j=1}^N\left\|h^ {L}_i-h ^ {L}_j \right\|_2^2 A^ {L}_ {i j}+\gamma\|A^{L}\|_F^2.
\end{equation}
Here, the first term is the regularization for the smoothness of brain region features, which is used to encourage adjacent nodes to have similar features. The second term is the constraint on the sparsity of the graph, which is achieved by penalizing the Frobenius norm~\cite{wang2022frobenius} of the graph.

We further introduce an encoder loss $\mathcal{L}_{\mathrm{E}}$ for multi-state GNN-based encoder following previous work~\cite{chen2023learnable}. The final polynomial loss function $\mathcal{L}=  \mathcal{L}_{CL} +\alpha \mathcal{L}_{\mathrm{GL}}+\beta \mathcal{L}_{\mathrm{E}}$ can be constructed to pre-train the entire framework, where $\alpha$ and $\beta$ are trade-off parameters.

In the fine-tuning stage of the model, our goal is to efficiently transfer the model that is pre-trained on a large-scale unlabeled dataset to different downstream diagnostic tasks. Owing to the powerful ability of brain connectivity learners to extract the intrinsic information of brain networks, it can be directly applied to new brain disorder diagnosis tasks without retraining. Thus, we only fine-tune the brain connectivity network encoder for each specific disease in the diagnosis tasks.
Finally, average pooling is adopted to aggregate brain network information from two different perspectives, and fully connected layers are used for the classification and diagnosis of brain disorders. The entire fine-tuning model uses cross-entropy as classification loss $\mathcal{L}_{\mathrm{L}}$ combined with encoder loss in polynomials for supervised training and prediction.

\section{Experiments}
\subsection{Experimental Settings}
The proposed method is implemented on a single NVIDIA GeForce GTX TITAN X 12GB GPU, utilizing the PyTorch package within the Python platform. To ensure a fair comparison, all rs-fMRI data are preprocessed following a standardized protocol and divided into 116 ROIs using the AAL atlas. For all graph neural network (GNN)-based methods, including the proposed multi-state brain network encoder, two graph neural network layers are utilized. The optimization process used the Adam optimizer with a learning rate of $1e-4$, a weight decay of $1e-4$, and $400$ training epochs. The model is trained with the following parameters: iterative layers set to 3 ($L=3$), head number set to 4 ($m=4$), trade-off parameter set to 0.01 ($\gamma=0.01$), and both $\alpha$ and $\lambda$ set to $0.001$, respectively. Four classical metrics are used to evaluate the performance of different methods, including accuracy (ACC), sensitivity (SEN), specificity (SPE), and area under the receiver operating characteristic curve (AUC).

\begin{table*}[!t]
\centering
\caption{Classification results ($\%$) in terms of ``mean(standard deviation)" achieved by different methods in cross-dataset brain disorder versus health control classification tasks. The proposed method shows statistically significant improvements (with p < 0.05) over all compared methods.}\label{tab:other}
\resizebox{\textwidth}{!}{
\begin{tabular}{c |c c cc|ccc c|cccc}
\hline
& \multicolumn{4}{c|}{ADNI} & \multicolumn{4}{c|}{ABIDE} & \multicolumn{4}{c}{ADHD}\\
\hline
Methods & ACC & SEN & SPE & AUC & ACC & SEN & SPE & AUC & ACC & SEN & SPE & AUC \\
\hline

FC+SVM & 60.3(2.21) & 59.9(2.44) & 60.8(2.68) & 60.7(2.61) & 58.1(2.19) & 57.3(1.25) & 60.8(1.33) & 59.4(2.04) & 55.7(1.82) & 56.8(1.34) & 52.9(1.68) & 55.6(1.29)   \\

EC+SVM &  60.5(2.09) & 62.4(2.34) & 58.6(2.04) & 61.5(2.21) & 59.6(1.98) & 60.5(1.51) & 57.9(1.08) & 59.9(1.87)& 55.0(1.63) & 56.2(1.86) & 53.7(1.31)  & 55.2(1.57)  \\

GCN & 70.1(2.91) & 70.7(2.26) & 68.5(2.20) & 70.3(2.59) & 67.8(2.09)  & 60.8(3.12) & 72.3(2.63) & 67.5(2.88) & 64.3(2.85) &  65.7(2.30) & 63.8(2.69) & 65.8(2.54)\\

BrainNetCNN &  71.0(2.55) & 72.6(2.91) & 70.1(3.41) & 72.6(2.74) & 69.4(3.08) & 66.9(3.25) & 70.8(2.87) & 69.6(2.93) & 64.8(2.71) & 63.6(2.16) & 66.1(2.90) & 66.3(2.51)  \\

BrainGB & 75.4(1.68) & 77.3(1.75) & 74.8(2.30) & 76.2(2.23) & 74.1(1.78) & 73.2(2.07) & 77.3(2.29) & 75.5(2.06) & 66.5(2.72) & 65.4(2.85) & 68.2(2.89) & 68.7(2.36) \\
\hline

BrainUSL & 78.6(2.04) & 76.1(1.51) & 81.5(2.73) & 80.9(1.98) & 77.0(1.46)  & 73.5(2.19) & 80.6(1.82) & 79.1(1.79) & 66.8(1.84) & 65.0(1.97) & 68.1(2.13) & 68.5(2.05) \\

BrainGSL & 80.3(2.50) & 77.4(2.31) & 82.7(2.81) & 82.1(2.48) &  78.0(2.35) & 75.3(2.07) & 84.6(2.10) & 79.8(2.17) & 68.6(2.29) & 76.4(2.53) & 60.2(2.55) & 69.8(2.37) \\

UCGL & 81.6(1.52) & 77.2(1.78) & 83.4(2.01) & 84.5(1.54) &  79.9(2.28) & 74.7(1.58) & 87.1(1.94) & 82.2(1.88) & 70.7(1.79) & 76.5(1.83) & 62.4(1.09) & 70.9(1.62) \\

SSGNN & 82.4(1.91) & 77.9(1.56) & 85.6(2.94) & 85.1(2.08) &  82.3(2.11) & 77.8(1.36) & 89.5(2.31) & 84.8(2.14) & 72.8(1.94) & 78.4(1.85) & 68.1(1.36) & 72.5(1.56) \\

\hline
Ours (w/o pre-training) & 78.2(1.69) & 72.8(1.86) & 79.3(2.03) & 79.8(1.91) & 77.9(2.23) & 74.0(2.69) & 82.9(2.35) & 78.6(2.14) & 67.2(2.08) & 74.7(2.31) & 61.9(1.86) & 69.3(2.02) \\

Ours & \textbf{86.2(1.75)} & \textbf{79.7(1.92)} & \textbf{87.9(1.67)} & \textbf{87.6(1.86)} & \textbf{84.2(1.37)} & \textbf{79.5(1.33)} & \textbf{89.8(1.21)} & \textbf{86.3(1.65)} & \textbf{76.8(1.63)} & \textbf{79.6(1.92)} & \textbf{68.5(2.24)} & \textbf{78.2(1.49)}\\
\hline
\end{tabular}
}
\end{table*}

\subsection{Comparative Methods}
We compare our proposed model with ten alternative methods, these methods include: 1) FC/EC+SVM~\cite{dyrba2015multimodal}: The brain connectivity matrices representing functional and effective connectivity are converted into feature vectors, and these feature vectors are then utilized as input for a gaussian kernel support vector machine. 2) GCN~\cite{kipf2016semi}: Functional connectivity matrices are fed into a graph convolution network, which incorporates two graph convolutional layers, one graph pooling layer, and one fully connected layer. 3) BrainNetCNN~\cite{kawahara2017brainnetcnn}: BrainNetCNN is composed of three types of filters, including two edge-to-edge layers, one edge-to-node layer, and one node-to-graph layer. 4) BrainGB~\cite{cui2022braingb}: The BrainGB framework offers a modular implementation encompassing a collection of general design patterns for developing GNN models specifically designed for brain network analysis. 5) BrainUSL~\cite{zhang2023brainusl}: The BrainUSL framework comprises a graph generation module that utilizes the inner product to optimize brain networks and Pearson correlation-based functional connectivity networks to compose paired samples to construct a contrastive learning framework. 6) BrainGSL~\cite{wen2023graph}: The BrainGSL includes a local topological-aware encoder that takes the partially visible nodes as input and a node-edge bi-decoder that reconstructs the masked edges using the representations of both the masked and visible nodes. 7) UCGL~\cite{wang2023unsupervised}: The UCGL is composed of a bi-level fMRI augmentation module based on slicing windows and a parallel feature extraction module. Each feature extraction module includes a spatio-temporal graph convolutional network and a multilayer perceptron predictor. 8) SSGNN~\cite{chen2024self}: The SSGNN utilizes the absolute value to generate brain networks and fine-tune the whole framework in the downstream tasks.
9) w/o pre-training: A variant of the proposed method, which has no pre-training stage.



Taking the experiment on ADNI as an example, two traditional methods (i.e., FC/EC+SVM), three deep learning methods (i.e., GCN, BrainNetCNN, and BrainGB), and the proposed w/o pre-training method are directly trained and tested on the ADNI dataset in a supervised manner.
Four unsupervised methods (i.e., BrainUSL, BrainGSL, UCGL, and SSGNN) and the proposed method are first pre-trained on the UKB dataset and then fine-tuned on the ADNI dataset. At the fine-tuning stage, we use 5-fold cross-validation and report the average and standard deviation of the testing results.


\subsection{Performance on Brain Disorder Classification}

We compare our proposed method with several recent studies in three common brain disorder diagnostic tasks. As shown in TABLE~\ref{tab:other}, it has the following observations: First, our method, along with graph-based deep learning methods, significantly outperforms the two traditional machine learning methods (i.e., FC/EC + SVM). These results suggest that graph-based deep learning methods, which can automatically mine discriminative brain structure information, are more effective in disease detection compared to traditional machine learning methods that rely on handcrafted fMRI features.
Second, the unsupervised pre-training methods (i.e., BrainUSL, BrainGSL, UCGL, SSGNN, and the proposed method) exhibit superiority over the supervised methods. For instance, in the ADNI dataset, the proposed method improves ACC by 8$\%$ and AUC by 7.8$\%$ compared to the best supervised method (i.e., w/o pre-training). This improvement may be attributed to unsupervised methods being capable of learning intrinsic information from a large amount of unlabeled data, which can aid in the analysis of new brain disorders. Third, our method significantly outperforms other methods in terms of all metrics. This improvement can be attributed to the optimal brain connectivity networks and representations through iterative jointly optimized learner and encoder.

\begin{table}[!t]
\centering

\caption{Comparison with state-of-the-art methods in five different classification tasks (i.e. NC vs. MCI, NC vs. ASD, and NC vs. ADHD) based on fMRI data. P: Positive Subjects; N: Negative Subjects. The best performances are in BOLD.}
\resizebox{\linewidth}{!}
{
\begin{tabular}{l l l l c c}
\hline

\multicolumn{1}{l}{\multirow{2}{*}{Task}}  & 
\multicolumn{1}{l}{\multirow{2}{*}{Method}}  & 
\multicolumn{1}{l}{\multirow{2}{*}{Year}}  & 
\multicolumn{1}{l}{\multirow{2}{*}{Subject (P/N)}} &  
\multicolumn{2}{c}{Performance (\%)} \\
\cline{5-6}
\multicolumn{1}{l}{} & \multicolumn{1}{l}{} & \multicolumn{1}{l}{} & \multicolumn{1}{l}{} & \multicolumn{1}{c}{ACC} & \multicolumn{1}{c}{AUC}\\ 
\hline

\multirow{4}*{NC vs. MCI}  & MAHGCN~\cite{liu2023hierarchical} & 2023 & 364/117 & 78.60 & 77.30 \\
~ & RandomFR~\cite{Liu2024tmi} & 2024 & 88/60 & 80.10 & 78.30 \\
~ & BIGFormer~\cite{10648828} & 2025 & 239/235 & 84.99 & 86.14 \\
~ & Ours & 2025 & 60/54 & {\textbf{86.21}} & {\textbf{87.59}} \\



\hline

\multirow{4}*{NC vs. ASD} & STCAL~\cite{liu2023spatial} & 2023 & 530/505 & 73.00 & 78.20 \\
~ & MSSTAN~\cite{10643532} & 2024 & 328/290 & 72.28 &  71.17 \\
~ & ATPGCN~\cite{10233896} & 2024 & 349/314 & 76.05 & 77.00 \\
~ & Ours & 2025 & 104/78  &  {\textbf{84.18}} & {\textbf{86.26}} \\

\hline

\multirow{4}*{NC vs. ADHD}  & MDCN~\cite{yang2023deep} & 2023 & 311/261 & 67.45 & 65.39 \\
~ & NSGA-III/S~\cite{9969465} & 2024 & 581/358 & 72.70 & 78.20 \\
~ & SCDA~\cite{fang2025source} & 2025 & 74/66 & 68.00 & 72.79 \\
~ & Ours & 2025 & 110/147 & {\textbf{76.77}} & {\textbf{78.23}} \\

\hline

\end{tabular}}
\label{table:sota}
\end{table}

\begin{table*}[!t]
	\centering
	\caption{Ablation study of the proposed method with different modules on cross-dataset classification tasks ($\%$).}
\resizebox{\textwidth}{!}
{
 \begin{tabular}{
  p{20pt}<{\centering} |
 p{20pt}<{\centering} p{20pt}<{\centering}
 |p{20pt}<{\centering} p{20pt}<{\centering}
 |p{20pt}<{\centering} p{20pt}<{\centering}
 |p{40pt}<{\centering} p{40pt}<{\centering}
 |p{40pt}<{\centering} p{40pt}<{\centering}
 |p{40pt}<{\centering} p{40pt}<{\centering}}
	
  \hline

\multicolumn{1}{c|}{\multirow{3}{*}{Model}} &
\multicolumn{2}{c|}{Learner} & \multicolumn{2}{c|}{Encoder} & \multicolumn{2}{c|}{Graph Loss} & \multicolumn{6}{c}{Metrics}
\\
\cline{2-13}
&  \multicolumn{1}{c}{\multirow{2}{*}{Adaptive}}
&  \multicolumn{1}{c|}{\multirow{2}{*}{Fixed}}
& \multicolumn{1}{c}{\multirow{2}{*}{Specific}}
& \multicolumn{1}{c|}{\multirow{2}{*}{Classical}}

&\multicolumn{1}{c}{\multirow{2}{*}{w/}}
& \multicolumn{1}{c|}{\multirow{2}{*}{w/o}}
& \multicolumn{2}{c|}{ADNI}
& \multicolumn{2}{c|}{ABIDE}
& \multicolumn{2}{c}{ADHD}

\\

\cline{8-13}
&&&&&&&ACC&AUC&ACC&AUC&ACC&AUC\\
\hline
1 & \multicolumn{1}{c}{$\surd$}  & \multicolumn{1}{c|}{} & \multicolumn{1}{c}{$\surd$}  & \multicolumn{1}{c|}{}  & \multicolumn{1}{c}{}  & \multicolumn{1}{c|}{$\surd$} & \multicolumn{1}{c}{56.7(2.98)}  & \multicolumn{1}{c|}{58.0(3.21)}  & \multicolumn{1}{c}{56.1(3.29)} & \multicolumn{1}{c|}{54.4(3.59)}   & \multicolumn{1}{c}{54.6(3.02)}  & \multicolumn{1}{c}{55.3(3.23)} 
\\
2 & \multicolumn{1}{c}{$\surd$}  & \multicolumn{1}{c|}{} & \multicolumn{1}{c}{}  & \multicolumn{1}{c|}{$\surd$}  & \multicolumn{1}{c}{$\surd$}  & \multicolumn{1}{c|}{} & \multicolumn{1}{c}{82.5(2.01)}  & \multicolumn{1}{c|}{83.7(2.28)} & \multicolumn{1}{c}{81.9(2.78)} & \multicolumn{1}{c|}{84.8(2.04)}   & \multicolumn{1}{c}{73.5(1.98)}  & \multicolumn{1}{c}{72.9(1.83)} 
\\

3& \multicolumn{1}{c}{$\surd$}  & \multicolumn{1}{c|}{} & \multicolumn{1}{c}{}  & \multicolumn{1}{c|}{$\surd$}  & \multicolumn{1}{c}{}  & \multicolumn{1}{c|}{$\surd$} & \multicolumn{1}{c}{55.3(2.94)}  & \multicolumn{1}{c|}{57.6(2.31)}  & \multicolumn{1}{c}{54.6(2.79)} & \multicolumn{1}{c|}{55.3(2.51)}   & \multicolumn{1}{c}{54.7(2.84)}  & \multicolumn{1}{c}{57.1(2.99)} 
\\
4& \multicolumn{1}{c}{}  & \multicolumn{1}{c|}{$\surd$} & \multicolumn{1}{c}{$\surd$}  & \multicolumn{1}{c|}{}  & \multicolumn{1}{c}{}  & \multicolumn{1}{c|}{$\surd$} & \multicolumn{1}{c}{80.8(2.53)}  & \multicolumn{1}{c|}{82.2(2.82)}  & \multicolumn{1}{c}{78.0(2.41)} & \multicolumn{1}{c|}{79.3(2.16)}   & \multicolumn{1}{c}{70.3(1.76)}  & \multicolumn{1}{c}{71.8(1.29)} 
\\


5& \multicolumn{1}{c}{}  & \multicolumn{1}{c|}{$\surd$}& \multicolumn{1}{c}{}  & \multicolumn{1}{c|}{$\surd$}  & \multicolumn{1}{c}{}  & \multicolumn{1}{c|}{$\surd$} & \multicolumn{1}{c}{77.5(1.94)}  & \multicolumn{1}{c|}{79.3(2.13)}  & \multicolumn{1}{c}{76.2(1.47)} & \multicolumn{1}{c|}{76.3(1.91)}   & \multicolumn{1}{c}{67.9(2.03)}  & \multicolumn{1}{c}{66.6(1.89)}   
\\
6 & \multicolumn{1}{c}{$\surd$}  & \multicolumn{1}{c|}{} & \multicolumn{1}{c}{$\surd$}  & \multicolumn{1}{c|}{}  & 
\multicolumn{1}{c}{$\surd$}  & \multicolumn{1}{c|}{} & \multicolumn{1}{c}{\textbf{86.2(1.75)}}  & \multicolumn{1}{c|}{\textbf{87.6(1.86)}}  & \multicolumn{1}{c}{\textbf{84.2(1.37)}} & \multicolumn{1}{c|}{\textbf{86.3(1.65)}}   & \multicolumn{1}{c}{\textbf{76.8(1.63)}}  & \multicolumn{1}{c}{\textbf{78.2(1.49)}}    
\\
       \hline
	\end{tabular}}
	\label{table:ablationstudy}
\end{table*}

\subsection{Comparison With State-of-the-Art}
We further compare our method with various state-of-the-art studies in the literature on three brain disorder diagnosis tasks using fMRI data. We present a summary of the State-of-the-Art (SOTA) methods published in TABLE~\ref{table:sota}. These methods include graph-based brain network analysis methods and non-graph-based methods. For each method, we provide the year of publication, the number of subjects included in the studies, as well as their respective accuracy and AUC values. It is important to note that the metric values presented in the table may vary depending on the dataset and experimental conditions.
It is obvious that our method outperforms existing SOTA methods in all classification tasks. For example, in the NC vs. ASD task, our method demonstrates significantly superior performance compared to other methods including graph-based methods and non-graph-based methods. This superiority may be attributed to the fact that the proposed method can learn fundamental brain network information from extensive data through pre-training, significantly enhancing its performance in downstream tasks.

\section{Discussion}
\subsection{Ablation Study}
In this experiment, we validate the effectiveness of three main modules in the framework: (1) brain connectivity network learner, (2) multi-state GNN-based encoder, and (3) graph loss.
Specifically, in the learner module, we compare the proposed learnable multi-head graph structure learner with predefined methods that construct fixed brain connectivity networks using Pearson correlation and transfer entropy during the data preprocessing stage. In the encoder module, we compare the multi-state encoder tailored for the multifunctional features of brain networks, with the classical graph convolutional network to evaluate the effectiveness of the specialized encoder. Finally, when the learner module adopts the learnable way, we compare the effects of with or without graph loss to investigate the impact of graph structure constraints on the model.

The results of all ablation studies are detailed in Table~\ref{table:ablationstudy}. By comparing model 6 with models 4 and 5, we find that the multi-head graph structure learner constructed in a learnable manner significantly enhances the model's performance on all tasks compared to predefined methods. Additionally, by comparing models 1 and 3 as well as models 4 and 5, we find that regardless of the variations in the initial graph structure, a specialized encoder tailored to the attributes of brain networks consistently improves the model's ability in brain network representation learning. Finally, by comparing model 1 with model 6 and model 2 with model 3, it is obvious that the absence of graph loss significantly reduces the model's performance. This effect may arise because the graphs generated in a learnable way without spare constraints are more susceptible to noise, complicating the analysis of brain networks.

\subsection{Strategy Analysis on Brain Connectivity Network Learner}
In the brain connectivity network learner, we design a cosine similarity metric function based on a multi-head mechanism and train the model through a joint iterative optimization method to obtain an optimal brain network learner. To verify the effectiveness of these strategies, we further conduct analytical experiments. Specifically, in addition to the cosine similarity utilized in this paper, we also compare other commonly used similarity metric functions, such as absolute value, inner product, and euclidean distance. We also analyze the scenario where the multi-head mechanism is not used. For joint iterative optimization, we set up two comparison strategies: a non-iterative method (i.e., with only one iteration) and a non-joint optimization method (i.e.,  only the encoder is iteratively optimized without updating the learner). 


As can be seen from Table~\ref{table:strategy}, it is evident that the proposed method is superior to all the alternative strategies in all tasks, and each independent strategy (i.e. cosine similarity, multi-head mechanism, and joint optimization) contributes to the efficacy of the final model. In addition, by comparing the results of single-head and non-iterative strategies, we find that they lead to an increase in the standard deviation of the model. This may be because the multi-head mechanism enhances the nonlinear modeling capabilities and the richness of feature expression by concurrently learning features from disparate subspaces, thereby significantly boosting the robustness of the model. Moreover, iterative optimization minimizes volatility during training by incrementally updating parameters, facilitating more stable convergence.

\begin{table}[!t]
    \centering
    \caption{Comparisons of ACC result in different Strategies ($\%$).}
    {
    \begin{tabular}{cccc}

    \hline
    Strategy  &  ADNI &  ABIDE &  ADHD \\
    \hline
    Absolute Value & 84.3(1.82) & 81.8(1.06) & 74.5(1.52) \\
    Inner Product & 85.9(1.97) & 83.2(1.73) & 75.9(1.77) \\
    Euclidean Distance & 84.7(1.69) & 82.4(1.54) & 76.1(1.91) \\
    \hline
    Single Head  & 84.5(2.44) & 83.5(2.34) & 74.9(2.21) \\
    \hline
    Non-Iterative  & 83.4(2.38) & 80.3(2.13) & 74.3(2.06) \\
    Non-Joint  & 85.1(2.01) & 82.9(1.90) & 76.1(1.88) \\
    \hline
    Proposed  & \textbf{86.2(1.75)} & \textbf{84.2(1.37)} & \textbf{76.8(1.63)}  \\
    \hline
    \end{tabular}}
    \label{table:strategy}
\end{table}


	


\subsection{Generalization Analysis}
\begin{figure}[!t]
	\begin{center}
		{\centering\includegraphics[width=0.8\linewidth]{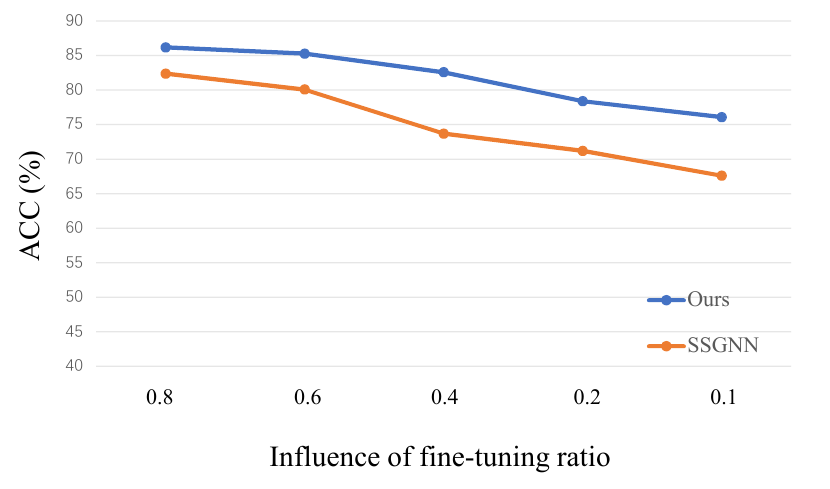}}
		\caption{Influence of fine-tuning ratio on the ADNI dataset, our proposed method in comparison with the second best method.}
	    \label{fintune}
	\end{center}
\end{figure}

We further evaluate the generalization ability of our proposed method by conducting experiments under varying fine-tuning ratios. Specifically, we fine-tune the pre-trained model on the ADNI dataset using different proportions of labeled samples: $\{0.8, 0.6, 0.4, 0.2, 0.1\}$. The performance of our method, in comparison with the second-best approach (i.e., SSGNN), is illustrated in Fig.~\ref{fintune}. As shown in the results, our method maintains competitive performance even with fewer fine-tuning samples. Moreover, the performance degradation is relatively slower as the fine-tuning ratio decreases, demonstrating the strong generalization capability of our model in cross-dataset brain disorder diagnosis.

\subsection{Parameter Sensitivity Analysis}
\begin{figure}[!t]
	\begin{center}
		{\centering\includegraphics[width=0.9\linewidth]{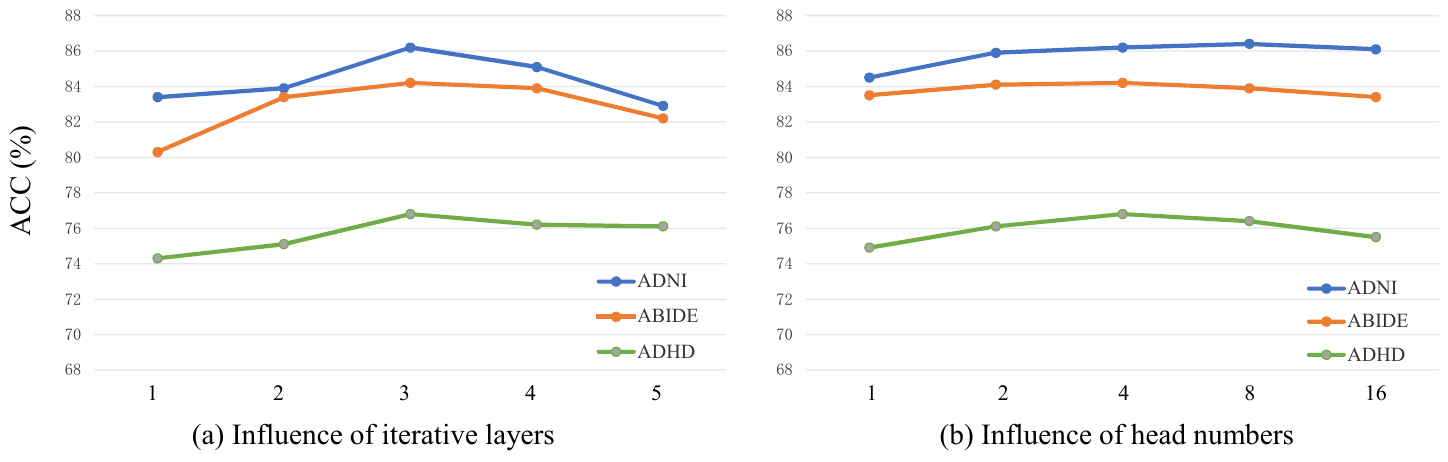}}
		\caption{Influence of two hyper-parameters on three different datasets.}
	    \label{parameter}
	\end{center}
\end{figure}

To investigate the impact of two hyper-parameters (i.e., $L$, and $m$), we perform grid search and report the ACC results of our proposed method in three different tasks.
As depicted in Fig.~\ref{parameter} (a), the accuracy of our method consistently increases with the number of iterative layers, reaching its peak at $L=3$ and subsequently displaying a significant decrease. This finding suggests that multiple iterations can effectively enhance the model's ability to extract feature information; however, excessively deep iterative layers may result in over-smoothing within the graph, subsequently diminishing the model's performance. Furthermore, Fig.~\ref{parameter} (b)  shows that our proposed method achieves its optimal performance at $m=4$ in both ABIDE and ADHD datasets, and tends to converge in the ADNI dataset. In summary, based on the model's performance, we ascertain the optimal combinations of parameters for each dataset and report the corresponding results.

\subsection{Brain Connectivity Network Analysis}
\begin{figure}[!t]
	\begin{center}
		{\centering\includegraphics[width=1.0\linewidth]{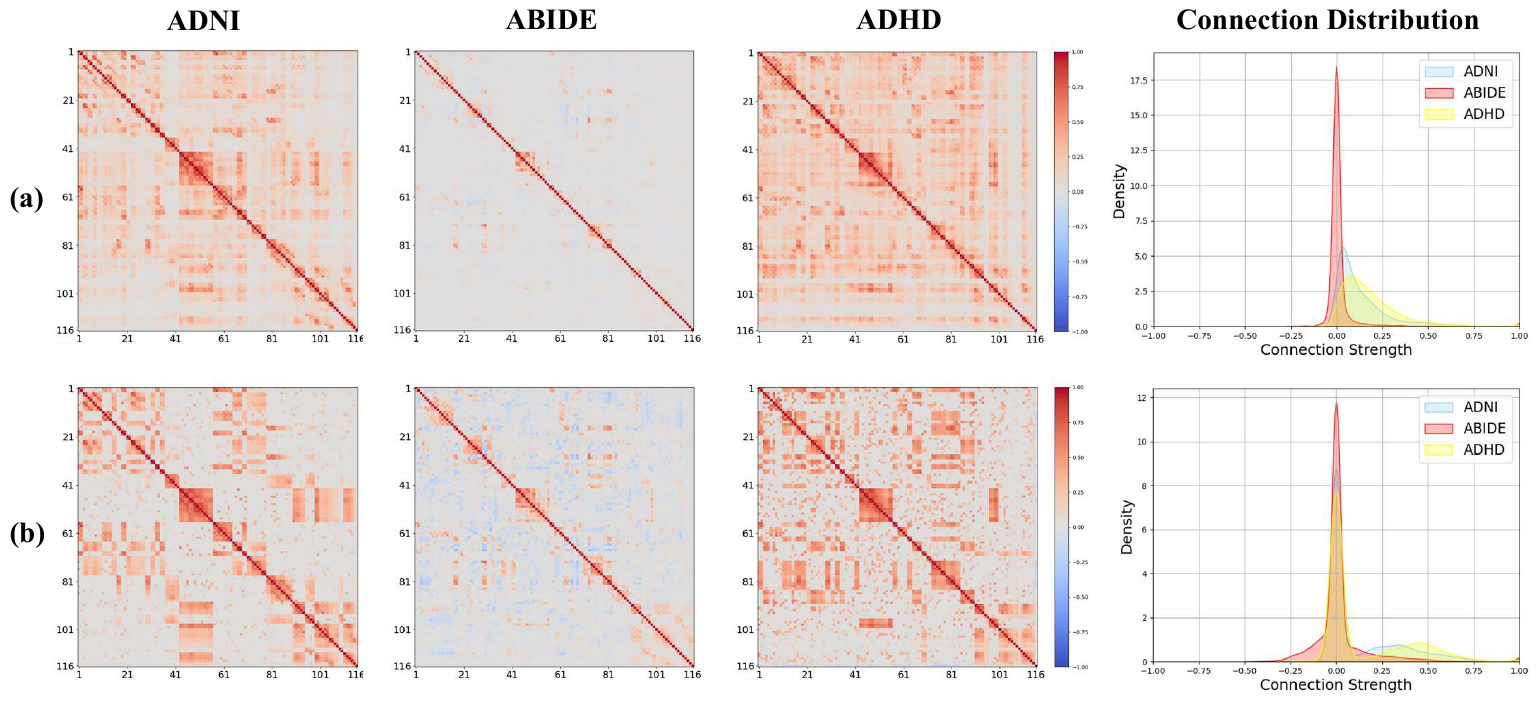}}
		\caption{Differences in functional connectivity patterns constructed by (a) predefined and (b) proposed method.}
	    \label{fc_compare}
	\end{center}
\end{figure}

To investigate the effectiveness of our proposed brain connectivity network learner, we conduct comparative experiments with traditional predefined approaches on three datasets. Specifically, we apply both the traditional predefined method (i.e., Fig.~\ref{fc_compare} (a)) and our proposed learnable approach (i.e., Fig.~\ref{fc_compare} (b)) to generate functional connectivity matrices for all samples within each dataset, and further visualizes the density distribution of these connectivity strengths.

As shown in Fig.~\ref{fc_compare}, it is obvious that compared with the traditional method, the proposed method calculates a clearer connection matrix structure with less noise (e.g., ADNI, and ADHD), and exhibits a more concentrated trend in the distribution of connection strength. Meanwhile,  more valuable information has been retained for ABIDE data that was originally too sparsely connected. This may be because the learnable method effectively removes irrelevant noise and emphasizes key functional connection patterns when calculating connection relationships. In addition, from the distribution curves of different datasets, the proposed method effectively reduces the distribution differences between different data while retaining their unique attributes.


\subsection{Visualization Analysis}
\begin{figure}[!t]
	\begin{center}
		{\centering\includegraphics[width=0.9\linewidth]{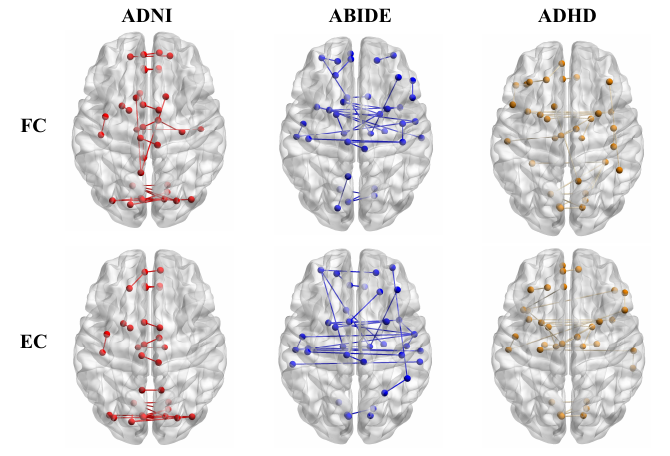}}
		\caption{Visualization of brain functional connectivity (FC) networks and effective connectivity (EC) networks generated by our method on three different datasets.}
	    \label{connections}
	\end{center}
\end{figure}

We further compare the performance of generated functional connectivity and effective connectivity on different datasets (ADNI, ABIDE, ADHD), in order to explore their complementarity in brain network analysis. In practice, we visualized the 30 most discriminative brain connections detected by our method in three datasets, as shown in Fig.~\ref{connections}.

The experimental results indicate that both functional connectivity and effective connectivity can discover common significant brain regions and connections. For example, in the ADNI dataset, both FC and EC reveal strong connections between bilateral frontal lobes, which are known to be among the core regions affected in the early stages of Alzheimer's disease~\cite{harwood2005frontal}. In the ADHD dataset, both types of connectivity appear sparse, suggesting a consistent capacity of FC and EC to reflect disease-related brain network abnormalities. In addition to this consistency, FC and EC also exhibit a degree of complementarity. For instance, in the ABIDE dataset, EC tends to capture more long-range connections, whereas FC highlights densely connected local regions. These findings suggest that our method of integrating FC and EC information can provide a more comprehensive characterization of brain connectivity networks, offering valuable insights for brain disorder diagnosis.


\section{Conclusion}
In this paper, we present an innovative framework for learning brain connectivity networks and diagnosing brain disorders. 
Our proposed method adopts a two-stage learning paradigm: first, pre-training is performed on unlabeled brain imaging data. We develop dual brain connectivity network learners to separately learn functional and effective connectivity patterns, and collaborate with a multi-state encoder to capture brain network features. We further design a joint iterative optimizing strategy to extract robust representation patterns across diseases
Subsequently, during the fine-tuning phase, the pre-trained network parameters are retained, simply adjust the brain connectivity network encoder for different disease tasks to quickly deploy. Our future plans include expanding the proposed method by incorporating inputs from diverse datasets and enhancing its applicability in real-world environments by reducing the reliance on fine-tuning target data.	

\bibliographystyle{model2-names.bst}\biboptions{authoryear}
\bibliography{refs}



\end{document}